\documentclass[prd,aps,twocolumn]{revtex4}
\usepackage{graphicx}

\font\smallrm=cmr8

\def \d {{\rm d}}

\def\boxit#1{\vbox{\hrule\hbox{\vrule\kern3pt
             \vbox{\kern3pt#1\kern3pt}\kern3pt\vrule}\hrule}} 

\begin{document}

\title{\bf Accelerating Kerr--Newman black holes in (anti-)de~Sitter space-time} 

\author{J. Podolsk\'y$^1$ and J. B. Griffiths$^2$}

\affiliation{ $^1$Institute of Theoretical Physics, Faculty of Mathematics and Physics, Charles University in Prague,
 V Hole\v{s}ovi\v{c}k\'ach 2, 18000 Prague 8, Czech Republic. \\
 $^2$Department of Mathematical Sciences, Loughborough University, 
 Loughborough,  Leics. LE11 3TU, U.K. }

\date{\today}

\begin{abstract}
\noindent
A class of exact solutions of the Einstein--Maxwell equations is presented which describes an accelerating and rotating charged black hole in an asymptotically de~Sitter or anti-de~Sitter universe. The metric is presented in a new and convenient form in which the meaning of the parameters is clearly identified, and from which the physical properties of the solution can readily be interpreted. 
\end{abstract}

\maketitle

\section{Introduction}

The Pleba\'nski--Demia\'nski \cite{PleDem76} family of solutions is known to include the case which describes an accelerating and rotating black hole. Significantly, the parameters contained in the metric include an arbitrary cosmological constant $\Lambda$ and electric and magnetic charge parameters. 
The solutions must therefore also contain particular cases which describe accelerating and rotating charged black holes in asymptotically de~Sitter or anti-de~Sitter universes. However, in the original form of the metric, these special cases are not clearly identified. Moreover, that form is not well suited for their physical interpretation. 
As these solutions are now being used for various new purposes, it is most important that they be better understood at the classical level. The purpose of the present paper is to contribute to such an understanding, particularly by expressing the metric in a form that is suitable for interpretation and in which the parameters involved have clear physical meanings.

For the non-rotating case, the Pleba\'nski--Demia\'nski family includes the $C$-metric whose analytic extension describes a causally separated pair of black holes which accelerate away from each other under the action of ``strings'' represented by conical singularities located along appropriate sections of the axis of symmetry. Hawking and Ross \cite{HawRoss95} have used this solution to describe the possible creation of a black hole pair by the breaking of a
cosmic string. The inclusion of a cosmological constant could alternatively be considered to supply the potential energy that is necessary for such a pair creation process. This has been analysed in \cite{ManRos95}--\cite{BooMan99}, where the cosmological constant was included in the traditional way. However, as shown in \cite{PodGri01}, it would have made more sense physically if $\Lambda$ had been inserted in an alternative way. This has been investigated in  \cite{DiaLem04,Dias04}. A particular case of the $C$-metric with $\Lambda<0$ has also been used for the costruction of a solution describing a black hole on the brane \cite{EmpHorMye00a,EmpHorMye00b}.

In addition, it was previously thought appropriate to use a coordinate freedom to remove the linear terms in the quartic functions which characterise the Pleba\'nski--Demia\'nski family of solutions. This was thought to remove the NUT parameter.
However, Hong and Teo \cite{HongTeo03} have recently shown that, at least for some special cases, the available freedom is much better used to simplify the roots of these functions. In the rotating case, they obtained a new solution \cite{HongTeo05} for an accelerating and rotating black hole which differs from what is usually called the ``spinning $C$-metric'' \cite{FarZim80b}--\cite{Pravdas02} in which the linear terms are set to zero. 
Surprisingly, it is the new solution of Hong and Teo which represents the NUT-free case, while the ``spinning $C$-metric'' retains NUT-like properties (i.e. part of the axis corresponds to a ``torsion'' singularity which is surrounded by a region containing closed timelike lines).

For the case in which $\Lambda=0$, a more general form of the metric was presented in \cite{GriPod05}. This has confirmed that the ``spinning $C$-metric'' does indeed possess an effective non-zero NUT parameter. It also introduces new parameters which explicitly describe the acceleration and rotation of the sources. 
In general, it covers the complete family of exact solutions which represent accelerating and rotating black holes with possible electromagnetic charges and an arbitrary NUT parameter, and describes the internal horizon and singularity structure as far as the associated acceleration horizon. The extension of this solution to boost-rotation-symmetric coordinates which cover both black holes is given in~\cite{GriPod06a}.

It may also be mentioned that the Hong--Teo paper \cite{HongTeo03}, for the non-rotating case with $\Lambda=0$, has already been extremely useful in leading to a better understanding of the higher-dimensional (rotating) black ring solution \cite{EmpRea02}. The improved factorizable structure introduced in \cite{HongTeo03} was explicitly used in \cite{ElvEmp03} and \cite{Emparan04}, and this new version has now became commonly employed in more recent investigations of black rings such as \cite{Harmark04}--\cite{CaDiYo05}.
To this form, we have here included both a Kerr-like rotation and an arbitrary cosmological constant. We have also adopted more physically motivated (Boyer--Lindquist-type) coordinates. It may be hoped that the solution described here may enable further solutions in higher dimensions and different backgrounds to be obtained and analysed.

The immediate purpose of the present paper, however, is to clarify the physical interpretation of the class of classical solutions in 3+1-dimensions when the cosmological constant is non-zero. We will concentrate here on the physically most significant case in which the space-time has no NUT-like properties.

\section{Accelerating and rotating charged black holes with $\Lambda\ne0$}

The Pleba\'nski--Demia\'nski metric covers a large family of solutions which includes that of an accelerating and rotating charged black hole. 
Among the various sub-families identified in \cite{GriPod06b}, we now present the following metric as the most convenient form for this particular case 
  \begin{equation}
  \begin{array}{l}
 {\displaystyle \d s^2={1\over\Omega^2}\bigg\{
 {Q\over\rho^2}\big[\d t- a\sin^2\theta\,\d\phi \big]^2 -{\rho^2\over Q}\,\d r^2 } \\[8pt]
 \hskip2.7pc {\displaystyle -{\rho^2\over P}\d\theta^2 -{P\over\rho^2}\sin^2\theta \big[ a\d t -(r^2+a^2)\d\phi \big]^2  
 \bigg\}, }
  \end{array}
  \label{lzeroMetric}
  \end{equation}
  where
  $$ \begin{array}{l}
  \Omega=1-\alpha\, r\cos\theta \,, \\[6pt]
  \rho^2 =r^2+a^2\cos^2\theta \,, \\[6pt]
  P= 1-2\alpha m\cos\theta  \\[3pt]
 \hskip3pc  +\Big(\alpha^2(a^2+e^2+g^2)+{1\over3}\Lambda a^2\Big)\cos^2\theta , \\[7pt]
 Q= \Big((a^2+e^2+g^2) -2mr +r^2\Big) (1-\alpha^2r^2)  \\[3pt]
 \hskip3pc  -{1\over3}\Lambda(a^2+r^2)r^2 .
  \end{array} $$ 
 This contains six arbitrary parameters $m$, $e$, $g$, $a$, $\alpha$ and $\Lambda$ which can each be varied independently. Besides the cosmological constant $\Lambda$, these parameters have distinct physical interpretations: $m$~is the mass of the black hole (at least in the non-accelerating limit), $e$ and $g$ are its electric and magnetic charges, $a$ measures its angular velocity, and $\alpha$ is its acceleration. (The possible NUT parameter $l$ has been put to zero here although, as shown in \cite{GriPod05}, the Pleba\'nski--Demia\'nski parameter $n$ must then be non-zero to avoid any NUT-like behaviour of the space-time.) The metric is accompanied by an electromagnetic field ${\hbox{F}=\d\hbox{A}}$, where the vector \hbox{potential is}
$$\hbox{A}={-er[\d t- a\sin^2\theta\,\d\phi \big]-g\cos\theta[ a\d t -(r^2+a^2)\d\phi]\over r^2+a^2\cos^2\theta}.$$

As advocated (for the case with ${\Lambda=0}$) by Hong and Teo \cite{HongTeo05} and clarified in \cite{GriPod05}, a coordinate freedom has been used in the derivation of (\ref{lzeroMetric}) to simplify the roots of the Pleba\'nski--Demia\'nski quartic function $\tilde P$ in this more general case. In particular, we can put 
$\tilde P(p)=(1-p^2)P(p)$. This enables us to put $p=\cos\theta$, so that we can work with conventional spherical polar coordinates (rather than complicated Jacobian elliptic functions) with $\theta\in[0,\pi]$ and $\phi$ a periodic coordinate whose precise period will be determined below. In accordance with this approach, the simpler function $P(\cos\theta)$ has been adopted above.

The only non-zero components of the curvature tensor relative to a natural null tetrad are
 $$  \begin{array}{l}
 {\displaystyle \Psi_2= \left(-m(1-i\alpha a)
+(e^2+g^2) {1+\alpha r\cos\theta\over r-ia\cos\theta} \right) } \\[8pt]
 \hskip3pc \times {\displaystyle 
 \left({1-\alpha r\cos\theta\over r+ia\cos\theta}\right)^3 }, \\[12pt]
 {\displaystyle \Phi_{11}= {1\over2}\,(e^2+g^2)\,{(1-\alpha r\cos\theta)^4\over(r^2+a^2\cos^2\theta)^2},
 \qquad \hbox{and} \quad \Lambda.}
  \end{array} $$ 
  These indicate the presence of a Kerr-like ring singularity at $r=0$, $\theta={\pi\over2}$. 
The vanishing of the conformal factor $\Omega$ corresponds to conformal infinity. Thus, we may take the range of $r$ as 
$ r\in(0,\alpha^{-1}\sec\theta )$ if $\theta<\pi/2$, and $ r\in(0,\infty)$ otherwise.
For $\theta\in({\pi\over2},\pi]$, the $r$ coordinate does not reach conformal infinity. In fact, an analytic extension through $r=\infty$ indicates \cite{GriPod06a} the presence of a second (mirror) region, as required for solutions expressed in boost-rotation-symmetric coordinates~\cite{BicakSchmidt89}.

As fully described in \cite{GriPod05} for the case with $\Lambda=0$, conical singularities generally occur on the axis. However, by specifying the range of $\phi$ appropriately, the singularity on one half of the axis can be removed. For example, that on $\theta=\pi$ is here removed by taking $\phi\in\big[0,2\pi(1+a_3-a_4)^{-1}\big)$, where $a_3=2\alpha m$ and 
$a_4=-\alpha^2(a^2+e^2+g^2)-{1\over3}\Lambda a^2$. In this case, the acceleration of the ``sources'' would be achieved by ``strings'' of deficit angle 
  \begin{equation}
 \delta_0 = {8\pi\,\alpha\,m \over
 1+2\alpha m+\alpha^2(a^2+e^2+g^2) +{1\over3}\Lambda a^2} , 
  \label{def0}
  \end{equation} 
 connecting them to infinity. Alternatively, the singularity on $\theta=0$ could be removed by taking $\phi\in\big[0,2\pi(1-a_3-a_4)^{-1}\big)$, and the acceleration would then be achieved by a ``strut'' between them in which the excess angle is
  \begin{equation}
 -\delta_\pi = {8\pi\,\alpha\,m \over
 1-2\alpha m+\alpha^2(a^2+e^2+g^2) +{1\over3}\Lambda a^2} . 
  \label{defpi}
  \end{equation} 
 Of course, the expressions (\ref{def0}) or (\ref{defpi}) are closely related to the forces in the string or strut respectively and these should be equal, at least according to Newtonian theory. However, it may be noticed that the deficit/excess angles are the same fractions of the range of the periodic coordinate in each case. Thus, they do correspond to identical expressions for the forces in the string or strut as expected, at least in the linear approximation.

In view of the fact that the complete space-time contains two accelerating black holes while the above coordinates only cover one of these, and also in view of the necessary presence of conical singularities, it may be observed that the space-time is not strictly asymptotic to de~Sitter or anti-de~Sitter space in all directions, except in the weak field limit.

Let us now note the following special cases in which one of the parameters $\alpha$, $\Lambda$ or $a$ vanishes respectively.

\subsection{The Kerr--Newman--(anti-)de~Sitter solution}

When $\alpha=0$, the metric (\ref{lzeroMetric}) reduces to that for the Kerr--Newman--(anti-)de~Sitter space-time. It can be expressed in standard Boyer--Lindquist-type coordinates \cite{GibHaw77} using the simple rescaling 
 \begin{equation}
 t=\bar t\,\Xi^{-1}, 
 \qquad \phi=\bar\phi\,\Xi^{-1},
 \label{KerrdSTrans}
 \end{equation}  
 where $\Xi=1+{1\over3}\Lambda a^2$. This puts the metric in the form 
 \begin{equation}
 \begin{array}{l}
 {\displaystyle \d s^2= {\Delta_r\over\Xi^2\rho^2}\Big[\d\bar
t-a\sin^2\theta\d\bar\phi\Big]^2 
 -{\rho^2\over \Delta_r}\,\d r^2 
 -{\rho^2\over\Delta_\theta}\,\d\theta^2 } \\[12pt]
 \hskip3pc {\displaystyle  
 -{\Delta_\theta\sin^2\theta\over\Xi^2\rho^2}\Big[a\d\bar t-(r^2+a^2)\d\bar\phi
\Big]^2 }, 
 \end{array}
 \label{KerrNdS}
 \end{equation} 
 where 
 \begin{equation}
 \begin{array}{l}
 \rho^2=r^2+a^2\cos^2\theta , \\[6pt]
 \Delta_r =(r^2+a^2)(1-{1\over3}\Lambda r^2) -2mr +(e^2+g^2) , \\[6pt]
 \Delta_\theta=1+{1\over3}\Lambda a^2\cos^2\theta . 
 \end{array}
 \end{equation} 
 Formally, there is no need to introduce the constant rescaling $\Xi$ in $t$ and $\phi$. However, this is included (at least for~$\phi$) so that the metric has a well-behaved axis at $\theta=0$ and $\theta=\pi$ with $\bar\phi\in[0,2\pi)$. It should also be noted that the metric (\ref{KerrNdS}) only retains Lorentzian signature for all $\theta\in[0,\pi]$ provided \ ${1\over3}\Lambda a^2>-1$.

\subsection{Accelerating and rotating black holes in a Minkowski background}

When $\Lambda=0$, the metric (\ref{lzeroMetric}) corresponds to that of Hong and Teo \cite{HongTeo05} (and described in detail in \cite{GriPod05}) which represents an accelerating and rotating pair of black holes without any NUT-like behaviour and in which the acceleration is identified as $\alpha$. In this case, if $m^2\ge a^2+e^2+g^2$, the expression for $Q$ factorises as 
 $$ Q = (r_--r)(r_+-r)(1-\alpha^2r^2), $$ 
  where 
  \begin{equation} 
  r_\pm = m\pm\sqrt{m^2-a^2-e^2-g^2}.
  \label{KerrNewman roots}
  \end{equation}
 The expressions for $r_\pm$ are identical to those for the locations of the outer and inner horizons of the non-accelerating Kerr--Newman black hole. However, in the present case, there is another horizon at $r=\alpha^{-1}$ which is already familiar in the context of the $C$-metric as an acceleration horizon.

In this case with $\Lambda=0$, the metric is equivalent to that given in equations (11)--(13) of Hong and Teo \cite{HongTeo05}, in which their coordinates $(t',x,y,\phi)$ are related to those used here by the transformation 
 $$ t'=-\alpha(t-a\phi), \qquad x=\cos\theta, \qquad y=1/(\alpha r), $$ 
 with $\phi$ unchanged, $A=\alpha$ and 
 $$ G(y)={1\over\alpha^2r^4}\,Q(r), \qquad G(x)=-\sin^2\theta\>P(\theta). $$ 
 However, the Boyer--Lindquist-type coordinates employed here seem to be physically more natural than the Pleba\'nski--Demia\'nski-type coordinates $x$ and~$y$. (The transformation in $t'$ is required for the existence of the axis at $\theta=0,\pi$.)

\subsection{The charged $C$-metric with a cosmological constant}

For the case in which $a=0$, the metric (\ref{lzeroMetric}) reduces to the simple diagonal form 
 $$ \begin{array}{l}
 {\displaystyle \d s^2={1\over(1-\alpha\, r\cos\theta)^2} \bigg( {Q\over r^2}\,\d t^2 
   -{r^2\over Q}\,\d r^2 } \\[12pt]
 \hskip5pc {\displaystyle  -{r^2\over P}\,\d\theta^2
   -P\,r^2\sin^2\theta\,\d\phi^2
  \bigg), }
 \end{array} $$ 
 where 
  $$ \begin{array}{l}
  P=1-2\alpha m\cos\theta +\alpha^2(e^2+g^2)\cos^2\theta, \\[6pt]
  Q=(e^2+g^2-2mr+r^2)(1-\alpha^2r^2)-{1\over3}\Lambda r^4.
  \end{array} $$ 
 This may be considered as a generalized and modified form for the charged $C$-metric that was introduced recently by Hong and Teo \cite{HongTeo03}. 
It describes a black hole of mass $m$ and electric and magnetic charges $e$ and $g$ which accelerates along the axis of symmetry under the action of forces represented by a topological (string-like) singularity, for which $\alpha$ is the acceleration, with an additional cosmological constant. When $\Lambda=0$ the black hole horizons (\ref{KerrNewman roots}) and the acceleration horizon at $r=\alpha^{-1}$ are clearly displayed. However, when $\Lambda\ne0$, the location of all horizons is modified.

Further properties of the charged $C$-metric in a de~Sitter or anti-de~Sitter background have been analysed in \cite{PodGri01} and \cite{DiaLem03a}--\cite{Krtous05}, using however different and less convenient forms of the metric to that presented above. When transformed to boost-rotation-symmetric coordinates, the new form above has a particularly simple structure, at least when $\Lambda=0$, as given in~\cite{GriPod06a}.

\section{Accelerating test particles and the nature of the new coordinates}

To elucidate the nature of the new coordinates introduced in the metric (\ref{lzeroMetric}), we now consider the weak field limit in which $m$, $a$, $e$ and $g$ are reduced to zero while $\alpha$ and $\Lambda$ remain arbitrary. The resulting metric is 
  \begin{equation} 
 \begin{array}{l}
 {\displaystyle \d s^2={1\over(1-\alpha\, r\cos\theta)^2} \Big[ 
 \big(1-(\alpha^2+{\textstyle{1\over3}}\Lambda)r^2\big)\,\d t^2 } \\[12pt] 
 \hskip2.5pc {\displaystyle -{\d r^2\over1-(\alpha^2+{\textstyle{1\over3}}\Lambda)r^2}  
 -r^2(\d\theta^2  +\sin^2\theta\,\d\phi^2) \Big] \!, }
 \end{array}
  \label{accdS} 
  \end{equation} 
 which reduces to the standard form of the Minkowski or (anti-)de~Sitter metric in static coordinates when $\alpha=0$.

Let us first observe that for $\Lambda=0$, the transformation 
\begin{equation} 
\begin{array}{l}
T={\displaystyle \ \ {\sqrt {\alpha^{-2}-r^2}\over 1-\alpha\,r\cos\theta}\,\sinh(\alpha\, t) }\,, \\[12pt]
Z={\displaystyle  \pm{\sqrt {\alpha^{-2}-r^2}\over 1-\alpha\,r\cos\theta}\,\cosh(\alpha\, t) }\,, \\[12pt]
R={\displaystyle {r\sin\theta\over1-\alpha\, r\cos\theta}} \,,
 \end{array}
  \label{accMink} 
 \end{equation} 
 leads to the standard form of the Minkowski line element 
 $$ \d s^2=\d T^2-\d Z^2-\d R^2-R^2\,\d\phi^2, $$ 
 confirming that all points with constant values of $r$, $\theta$ and $\phi$ are in \emph{uniform acceleration} in the positive or negative $Z$-direction relative to the Minkowski background. In particular, a test particle located at the origin $r=0$ of the new coordinates has acceleration given exactly by $\alpha$ as it moves along either of the trajectories \ 
$T=\alpha^{-1}\sinh(\alpha\, t)$, \ 
\hbox{$Z=\pm\alpha^{-1}\cosh(\alpha\, t)$}, \ $R=0$.

Similarly, when $\Lambda\ne0$, the metric (\ref{accdS}) just describes a de~Sitter or anti-de~Sitter universe but expressed in new accelerating coordinates. These space-times can be represented as the four-dimensional hyperboloid 
 $$ {Z_0}^2 -{Z_1}^2 -{Z_2}^2 -{Z_3}^2 -\epsilon{Z_4}^2  =-3/\Lambda\,, $$ 
 in the flat five-dimensional space 
  $$ \d s^2 ={\d Z_0}^2 -{\d Z_1}^2 -{\d Z_2}^2 -{\d Z_3}^2  
 -\epsilon{\d Z_4}^2\,,  $$ 
 where $\epsilon=\hbox{sign}\,\Lambda$.

Provided $\alpha^2+{1\over3}\Lambda>0$, the metric (\ref{accdS}) can be expressed in this notation by the parametrization 
 \begin{equation} 
 \begin{array}{l}
 Z_0=\ \ {\displaystyle {\sqrt{(\alpha^2+{1\over3}\Lambda)^{-1}-r^2} 
 \over 1-\alpha\,r\cos\theta} }\,\sinh(\sqrt{\alpha^2+{1\over3}\Lambda}\>t) \,, \\[12pt]
 Z_1= \pm{\displaystyle {\sqrt{(\alpha^2+{1\over3}\Lambda)^{-1}-r^2} 
 \over 1-\alpha\,r\cos\theta} }\,\cosh(\sqrt{\alpha^2+{1\over3}\Lambda}\>t) \,, \\[12pt]
 Z_2= {\displaystyle {r\sin\theta\sin\phi \over 1-\alpha\,r\cos\theta} } \,,\\[12pt]
 Z_3= {\displaystyle {r\sin\theta\cos\phi \over 1-\alpha\,r\cos\theta} } \,,\\[12pt]
 Z_4= {\displaystyle {\alpha-(\alpha^2+{1\over3}\Lambda)\,r\cos\theta
 \over\sqrt{{1\over3}|\Lambda|}\,\sqrt{\alpha^2+{1\over3}\Lambda}\,
 (1-\alpha\,r\cos\theta)} } \,.
 \end{array}
  \label{acchyperbol} 
 \end{equation} 
 (Notice that, in this case, a possibility exists to perform either of the limits $\alpha\to0$ or $\Lambda\to 0$.) 
The trajectory of a test particle located at $r=0$ is given by 
\begin{equation} 
 \begin{array}{l}
 Z_0=\ \ {(\alpha^2+{1\over3}\Lambda)^{-1/2}}\,\sinh(\sqrt{\alpha^2+{1\over3}\Lambda}\>t) \,, \\[12pt]
 Z_1= \pm{(\alpha^2+{1\over3}\Lambda)^{-1/2}}\,\cosh(\sqrt{\alpha^2+{1\over3}\Lambda}\>t) \,,
 \end{array}
  \label{orighyperbol} 
 \end{equation} 
$Z_2=Z_3=0$, $Z_4=\hbox{const.}$ \ This actually represents the trajectories of a pair of uniformly accelerated particles in a de~Sitter or anti-de~Sitter space-time (see e.g. \cite{PodGri01,DiaLem03a,DiaLem03b,BicKrt05}).

In the alternative case of a test particle with small acceleration in an anti-de~Sitter universe, for which ${\alpha^2+{1\over3}\Lambda<0}$, the metric (\ref{accdS}) corresponds to the parametrization
 \begin{equation} 
 \begin{array}{l}
 Z_0= {\displaystyle {\sqrt{r^2-(\alpha^2+{1\over3}\Lambda)^{-1}} 
 \over 1-\alpha\,r\cos\theta} }\,\sin(\sqrt{-(\alpha^2+{1\over3}\Lambda)}\>t) \,, \\[12pt]
 Z_4= {\displaystyle {\sqrt{r^2-(\alpha^2+{1\over3}\Lambda)^{-1}} 
 \over 1-\alpha\,r\cos\theta} }\,\cos(\sqrt{-(\alpha^2+{1\over3}\Lambda)}\>t) \,, \\[12pt]
 Z_1= {\displaystyle {\alpha-(\alpha^2+{1\over3}\Lambda)\,r\cos\theta
 \over\sqrt{{1\over3}|\Lambda|}\,\sqrt{-(\alpha^2+{1\over3}\Lambda)}\,
 (1-\alpha\,r\cos\theta)} } \,,
 \end{array}
  \label{acchyperbolsingle} 
 \end{equation} 
 with $Z_2, Z_3$ as in (\ref{acchyperbol}). In this case, the trajectory ${r=0}$ represents the motion of a single uniformly accelerated test particle in an anti-de~Sitter universe \cite{Pod02,DiaLem03a,Krtous05}.

In fact, \emph{any} world-line ${x^\mu(\tau)=(t(\tau),r_0,\theta_0,\phi_0)}$ in the space-time (\ref{accdS}), where ${r_0,\theta_0,\phi_0}$ are constants and $\tau$ is the proper time, represents the motion of a uniformly accelerated test particle. Its 4-velocity is ${U^\mu=(\dot{t},0,0,0)}$, $\dot{t}=1/\sqrt{g_{tt}(r_0,\theta_0)}=\hbox{const.}$, and the 4-acceleration has constant components 
$${A^\mu\equiv U^\mu_{\ ;\nu}U^\nu=\Big(0,{-g_{tt,r}\over2g_{tt}g_{rr}},{-g_{tt,\theta}\over2g_{tt}g_{\theta\theta}},0\Big)}\,.$$
Since ${A^\mu U_\mu=0}$, it is a spatial vector in the instantaneous rest frame orthogonal to the 4-velocity, and its constant magnitude is   
 \begin{equation} 
A^2\equiv -A^\mu A_\mu=(\alpha^2+{\textstyle{1\over3}}\Lambda){(1-\alpha\, r_0\cos\theta_0)^2\over 1-(\alpha^2+{1\over3}\Lambda)\,r_0^2}-{\textstyle{1\over3}}\Lambda\,.
  \label{acceleration}
 \end{equation}
In particular, the uniform acceleration of a test particle located at the origin ${r=0}$ is given exactly by ${A=\alpha}$, independently of $\theta_0,\phi_0$ or $\Lambda$. For this reason, we may conclude that the new form of the line element (\ref{lzeroMetric}) may be interpreted as using most convenient accelerated coordinates in a Minkowski or (anti-)de~Sitter background~(\ref{accdS}). (Of course, the acceleration is only that of a real physical particle when $m$ is non-zero and there exists a physical cause that can be modelled by (\ref{def0}) or  (\ref{defpi}).)

\section{Conclusion}

The metric (\ref{lzeroMetric}) is presented here as the most convenient form with which to analyse the properties of a rotating and accelerating, possibly charged, black hole in an asymptotically de~Sitter or anti-de~Sitter space-time. 
In particular, the parameters employed all possess an explicit physical interpretation.

This form of the metric nicely represents the horizon and singularity structure of the solution. It covers the space-time from the singularity, through the inner and outer black hole horizons, through the exterior region, and even through the acceleration horizon. It also nicely describes the conical singularity that is required to produce the acceleration. However, it it does not represent the complete analytical extension of the space-time, either through the black hole horizons or beyond the acceleration horizon. For such extensions, either a Kruskal--Szekeres-like transformation or a transformation to boost-rotation-symmetric coordinates is required respectively. Such extensions would reveal multiple possible sources inside the black hole horizon and mirror, causally separated sources beyond the acceleration horizon.

Let us finally note that the metric (\ref{lzeroMetric}) has clear limits both when $\alpha=0$ and when $\Lambda=0$, so that it is actually a better representation of an accelerating charged black hole in the above backgrounds than that given previously in \cite{PodGri01} (to which it is related by the rescaling ${t\to t\sqrt{1+{3\over\Lambda}\alpha^2}}$, 
${r\to- r\sqrt{1+{3\over\Lambda}\alpha^2}}\,$). In addition, the metric functions depend  on (powers of) $r$ and $\cos\theta$ only. More importantly, a non-vanishing Kerr-like rotation is now also included.

\section*{Acknowledgements}

The authors are grateful to the referees for some helpful comments on an earlier draft of this work, which was partly supported by grants from the EPSRC and (JP) by GACR 202/06/0041.

\end{document}